\documentclass{ws-procs975x65}
\usepackage{amsmath}
\usepackage{amssymb}

\begin{document}

\title{INFERRING THE NEUTRON STAR EQUATION OF STATE\\
FROM BINARY INSPIRAL WAVEFORMS}

\author{CHARALAMPOS MARKAKIS$^1$, JOCELYN S. READ$^2$, 
MASARU SHIBATA$^3$,\\ K\=OJI URY\=U$^4$,
JOLIEN D. E. CREIGHTON$^1$ and JOHN L. FRIEDMAN$^1$}

\address{$^1$Department of Physics, University of Wisconsin--Milwaukee\\
PO Box 413, Milwaukee, WI 53201, USA}
\address{$^2$Max-Planck-Institut f\"ur Gravitationsphysik, Albert-Einstein-Institut\\ Golm, Germany}
\address{$^3$Yukawa Institute for Theoretical Physics, Kyoto University\\  Kyoto 606-8502, Japan}
\address{$^4$Department of Physics, University of the Ryukyus\\ 1 Senbaru, Nishihara, Okinawa 903-0213, Japan}

\begin{abstract}
The properties of neutron star matter above nuclear density are not precisely known. Gravitational waves emitted from binary neutron stars during their late stages of inspiral and merger contain imprints of the neutron-star equation of state. Measuring departures from the point-particle limit of the late inspiral waveform allows one to measure properties of the equation of state via gravitational wave observations. 
This and a companion talk by J. S. Read reports a comparison
of  numerical waveforms from simulations of inspiraling
neutron-star binaries, computed for equations of state
with varying stiffness. 
We calculate the signal strength of the difference between waveforms for various commissioned and proposed interferometric gravitational wave detectors and show that observations at frequencies around 1 kHz will be able to measure a compactness parameter and constrain the possible neutron-star equations of state. 
\end{abstract}


\bodymatter\bigskip


In  simulations of the late inspiral and merger of binary neutron-star systems, one typically
specifies an equation of state (EOS) for the matter, performs a numerical evolution and extracts the gravitational
waveforms produced in the inspiral. In this talk we report work on the inverse problem:
if gravitational waves from an inspiraling neutron-star binary are observed, can they be used to infer the bulk
properties of neutron star matter and, if so, with what accuracy? To answer this question, we performed a number of
simulations \cite{PhysRevD.79.124033,Markakisetal2009,Read2008}, using the evolution and initial data codes of Shibata and Ury\=u,
while systematically varying the stiffness of a parameterized EOS. This parameterized EOS was previously developed in Refs. \refcite{PhysRevD.79.124032,Read2008} and is of piecewise polytropic form, $p(\rho)=K_i\rho^{\Gamma_i}$ in a set of three intervals $\rho_{i-1}\leqslant\rho\leqslant\rho_i$
in rest-mass density, with the constants $K_i$ determined by requiring continuity on each dividing $\rho_i$ and the energy density determined by the first law of thermodynamics. As described in Refs. \refcite{PhysRevD.79.124032,Read2008}, the nonpolytropic  EOS of the crust ($0\leqslant\rho\leqslant\rho_0$) as well as the dividing densities $\rho_1, \rho_2$
are fixed 
while the parameters 
$\{p_1\equiv p(\rho_1), \Gamma_1, \Gamma_2, \Gamma_3\}$
are generally varied. In this first set of simulations we set $\Gamma_1=\Gamma_2=\Gamma_3=3$ and change the EOS\ stiffness  by  varying only $p_1$, while  keeping the  Schwarzschild mass of each neutron star fixed at $1.35\,M_\odot$. The choice of EOS parameter varied in this work is motivated by the fact that neutron-star radius is closely tied to the pressure at density not far above nuclear equilibrium density \cite{LattimerPrakash2001}.
Variation of the adiabatic exponents is the scope of a next… set of simulations.


We compared the gravitational waveforms from the  simulations  to point-particle waveforms 
(see, for example, Ref. \refcite{BoyleEtal2007})
and calculated the signal strength of the difference in waveforms using the sensitivity curves of commissioned and proposed gravitational wave detectors. We find that, as the stars approach their final plunge and merger, the gravitational phase accumulates more rapidly for larger values of $p_1$ or smaller values of the neutron-star compactness (the ratio of the neutron-star mass to its radius).
The waveform analysis indicates that realistic EOS will result in waveforms that are distinguishable from point-particle inspiral at an effective distance (the distance to an optimally oriented… and located system that would produce an equivalent waveform amplitude) of $D_0 = 100~ \rm{Mpc}$ or less with   gravitational wave detectors with the sensitivity of broadband Advanced LIGO.
We further estimate that observations of this sensitivity will be able to constrain $p_1$  for a source at effective distance $D$ with an accuracy of
$\delta p_1/p_{1}\sim 0.2 D/D_0$. 
Related estimates of radius measurability show that such observations can determine the radius to an accuracy of 
$\delta R\sim 1 ~ \rm{km}$~$D/D_0$.
These first estimates neglect other  details of internal structure  which are expected to give smaller tidal effect corrections. This is the subject of work underway, which involves improving the accuracy of the estimates with variation of the adiabatic exponents, determination of surfaces in the equation of state (EOS) parameter space associated with a given departure from the waveform of point-particle inspiral
 and numerical simulation of more orbits in the late inspiral.   Also, the  results mentioned above do not take into account multiple detectors, parameter correlation, or multiple observations. The latter possibility is briefly discussed below.

In the  calculations  mentioned above we estimated the error $\sigma_0$ in measuring an EOS parameter (such as $p_1$ or, more precisely, a related parameter that labels surfaces of constant departure from point-particle inspiral) from observation of one event at a reference effective distance $D_0=100~\rm{Mpc}$. Here we wish to estimate the effect of multiple observations  on the measurement accuracy. 
If $N_i$ identical events, each with measurement uncertainty $\sigma_i$,
occurred at the same effective distance $D_i$, then the overall uncertainty (standard error) of the combined measurement would be $\sigma_i/ \sqrt{N_{i}}$.
However,  events do not  occur at the same effective distance.
Instead, we shall assume that events are homogeneously distributed
in a sphere of effective radius $D_{\max}\simeq300~\rm{Mpc}$.
(A uniform probability distribution of events in space is also uniform in effective space.) We divide this sphere into $I$  shells of effective distance $D_i=D_{\max}(i-\frac{1}{2})/I$ with $i=1,...,I$
and assume that detections will only be counted for sources with effective distance smaller than $D_{\max}$.
Because uncertainty scales linearly with effective distance, we have 
$\sigma_i=\sigma_0 D_i/D_0$.
  Combining measurements at different distances will then result in an overall uncertainty $\sigma$ given by
\begin{equation} \label{eq1}
\frac{1}{\sigma^2}=\sum_{i=1}^{I}\frac{N_i}{\sigma_i^2}=\frac{D_0^2}{\sigma_0^2}
\sum_{i=1}^{I} \frac{N_i}{D_i^2}
\end{equation}
where 
$N_i$ is the number of events in the $i$-th shell. Note that the $N_i$ are random variables, so 
$\sigma$ is also a random variable with some probability distribution. The total number of events in a given year,
\begin{equation} \label{eq2}
N=\sum_{i=1}^{I} N_i,
\end{equation}
is itself a random variable and is Poisson-distributed around the rate of events $\mathcal{R}\equiv\langle N\rangle$ (average number of events per year).
The probability distribution function $\mathcal{P}(\sigma|\mathcal{R})$ of $\sigma$ given $\mathcal{R}$, and its moments $\langle\sigma\rangle$,~$\langle\sigma^2\rangle$ etc., are more easily computed via Monte-Carlo simulation rather than analytically. However, an analytical estimate for the moment
$\langle\sigma^{-2}\rangle$
can be  obtained by  assuming a constant density $n$ of events per year per unit effective volume and converting the sums \eqref{eq2} and \eqref{eq1} to integrals:
\begin{eqnarray} \label{eq3}
\langle N\rangle&=&\int_0^{D_{\max}}4\pi n \, D^2
dD=\frac{4\pi n}{3} D_{\max}^3\\
\left\langle \frac{1}{\sigma^2} \right\rangle&=&\frac{D_0^2}{\sigma_0^2}\int_0^{D_{\max}}\frac{1}{D^2}\,4\pi n \, D^2
dD=\frac{D_0^2}{\sigma_0^2}4\pi n \,D_{\max} \label{eq4}
\end{eqnarray}
Eliminating $n$ from eqs. \eqref{eq3} and \eqref{eq4}  yields
the simple formula\begin{equation} \label{eq5}
\langle \sigma^{-2}\rangle^{-1/2}=\sigma_0\frac{D_{\max}}{D_0}(3\mathcal{R})^{-1/2}
\end{equation}
Our Monte-Carlo simulations confirm that, while  $\langle\sigma\rangle$,~$\langle\sigma^2\rangle^{1/2}$ and other moments do not in general scale proportionately to $\mathcal{R}^{-1/2}$ for fixed $D_{\max}$, the moment $\langle \sigma^{-2}\rangle^{-1/2}$ scales exactly as dictated by eq. \eqref{eq5}.
This formula indicates that, for example, three events randomly distributed in a sphere of effective radius $D_{\max}=3D_0$
give the same ``average'' uncertainty $\langle \sigma^{-2}\rangle^{-1/2}=\sigma_0$ as one event at effective distance $D_0$.
Although the above calculation was done for measurement of a single parameter, it can be straightforwardly generalized for more parameters, 
by replacing the uncertainty $\sigma^2$ with a  Fisher information matrix.
As noted above, we have restricted consideration to neutron stars with a single fixed mass. Independent variation of the mass of each companion is the subject of future work.
 \\
\textit{Acknowledgements:} This work was supported in part by NSF grants PHY-0503366, PHY-0701817 and PHY-0200852,
the Greek State Scholarships Foundation 
and JSPS Grants
 20540275 and 19540263.
 CM gratefully acknowledges travel support by NSF\   grant PHY-0919134
and assistance with the manuscript by M. Bakopoulou.

\bibliographystyle{ws-procs975x65}

\begin{thebibliography}{1}

\bibitem{PhysRevD.79.124033}
J.~S. Read \textit{et al.},
{\em Phys. Rev. D} {\bf 79}, p. 124033 (2009).

\bibitem{Markakisetal2009}
C.~Markakis \textit{et al.},
  {\em Journal of Physics: Conference Series} {\bf
  189}, p. 012024 (2009).

\bibitem{Read2008}
J.~S. Read,
 PhD thesis, University of Wisconsin - Milwaukee
  (WI, USA, 2008).

\bibitem{PhysRevD.79.124032}
J.~S. Read \textit{et al.},
 {\em Phys. Rev. D}  {\bf 79}, p. 124032 (2009).

\bibitem{LattimerPrakash2001}
J.~M. Lattimer and M.~Prakash, {\em The Astrophysical Journal} {\bf 550}, 426
  (2001).

\bibitem{BoyleEtal2007}
M.~Boyle \textit{et al.},
  {\em Phys. Rev. D} {\bf 76}, p.
  124038 (2007).


\end{thebibliography}


\begin{thebibliography}{1}

\bibitem{jarl88}
C.~Jarlskog, {\em CP {V}iolation} (World Scientific, Singapore, 1988).

\bibitem{lamp94}
L.~Lamport, {\em \LaTeX, A Document Preparation System}, 2nd edn.
  (Addison-Wesley, Reading, MA, 1994).

\bibitem{ams04}
\AmS, {\em \AmS-\LaTeX{} Version 2 User's Guide} (American Mathematical
  Society, Providence, 2004).
\newblock \url{http://www.ams.org/tex/amslatex.html}.

\end{thebibliography}

\end{document}